\documentclass[twocolumn,showpacs,preprintnumbers,amsmath,amssymb,prl,superscriptaddress]{revtex4}
\usepackage{graphicx}

\begin{document}

\title{Nonlinear screening of charge impurities in graphene}
\author{M. I. Katsnelson}
\email{M.Katsnelson@science.ru.nl} \affiliation{Institute for
Molecules and Materials, Radboud University Nijmegen, 6525 ED
Nijmegen, The Netherlands}


\begin{abstract}
It is shown that a ``vacuum polarization'' induced by Coulomb
potential in graphene leads to a strong suppression of electric
charges even for undoped case (no charge carriers). A standard
linear response theory is therefore not applicable to describe the
screening of charge impurities in graphene. In particular, it
overestimates essentially the contributions of charge impurities
into the resistivity of graphene.
\end{abstract}

\maketitle

\affiliation{Institute for Molecules and Materials, Radboud
University Nijmegen, 6525 ED Nijmegen, The Netherlands}

Graphene is a name given to an atomic layer of carbon atoms packed
into a hexagonal two-dimensional lattice. This term is widely used
to describe the crystal structure and properties of graphite
(which consists from graphene layers relatively loosely stacked on
top of each other), carbon nanutubes, and large fullerenes. Very
recently, a way has been found to prepare free-standing
graphene\cite{kostya0,kostya1}, that is, real two-dimensional
crystal (in contrast with numerous \textit{quasi}-two-dimensional
systems known before). The graphene turns out to be a gapless
semiconductor with a very high electron mobility which makes it a
perspective material, e.g., for ballistic field-effect
transistor\cite{kostya1}. It has been shown\cite{kostya2,kim} that
the charge carriers in graphene are massless Dirac fermions with
effective ``velocity of light'' of order of 10$^{6}$ ms$^{-1}$.
Due to this unusual electronic structure graphene demonstrates
exotic transport properties, such as a new kind of the integer
quantum Hall effect with half-integer quantization of the Hall
conductivity\cite{kostya2,kim,gus,per,cas,lee}, or finite
conductivity in the limit of zero charge-carrier concentration
\cite{kostya2,ktsn,been,zieg}.

One of peculiar transport properties of graphene, a mobility which
is almost independent on the charge carrier
concentration\cite{kostya2}, was explained in
Refs.\onlinecite{nomura,ando} as a result of electron scattering
by charge impurities. However, a linear-response theory was used
to take into account screening effects. Rigorously speaking, this
theory can be applied only assuming that the impurity potential is
small in comparison with the Fermi energy; however, even in
semiconductors where this condition can be, in general, broken
this theory can be normally used and gives reasonable results (for
the case of two-dimensional electron gas, see for review
Ref.\onlinecite{rmp}).

In this paper we calculate nonlinear screening of charge
impurities in graphene taking into account a ``vacuum
polarization'' effect in a region of strong potential.

A general nonlinear theory of screening in the system of
interacting particles can be formulated in a framework of the
density functional approach\cite{kohn}. In this theory a total
potential $V\left( \mathbf{r}\right)$ acting on electrons equals
\begin{equation}
V\left( \mathbf{r}\right) =V_{0}\left( \mathbf{r}\right)
+V_{ind}\left( \mathbf{r}\right)   \label{total}
\end{equation}
where $V_{0}\left( \mathbf{r}\right)$ is an external potential and
$V_{ind}\left( \mathbf{r}\right)$ is a potential induced by a
redistribution of electron density:
\begin{equation}
V_{ind}\left( \mathbf{r}\right) =\frac{e^{2}}{\epsilon }\int d\mathbf{r}%
^{\prime }\frac{n\left( \mathbf{r}^{\prime }\right)
-\overline{n}}{\left| \mathbf{r-r}^{\prime }\right| }+V_{xc}\left(
\mathbf{r}\right),  \label{DF}
\end{equation}
where the first term is the Hartree potential and the second one
is the exchange-correlation potential. We will consider here
explicitly only a redistribution of charge carriers in the
external impurity potential
\begin{equation}
V_{0}\left( \mathbf{r}\right) =\frac{Ze^{2}}{\epsilon r}
\label{impur}.
\end{equation}
taking into account contributions of crystal lattice potential and
of electrons in completely filled bands via dielectric constant
$\epsilon$ and compensated homogeneous charge density
$-e\overline{n}$; for the case of graphene on quartz one should
choose\cite{ando} $\epsilon \simeq 2.4-2.5$. Here $Z$ is the
dimensionless impurity charge (to be specific, we will assume
$Z>0$; it can be easily demonstrated that, actually, in our final
expressions $Z$ should be just replaced by $\left| Z\right| $).
This kind of approach is valid at a space scale much larger than a
lattice constant; in all other aspects, it is formally exact until
we specify the expressions for $V_{xc}$ and $n\left[V
\left(\mathbf{r} \right) \right]$.

A dimensionless coupling constant $\alpha = \frac{e^2}{\epsilon
\hbar v_F}$ (where $v_F \simeq 10^{6}ms^{-1}$ is the Fermi
velocity in graphene) determining the strength of interelectron
interactions is of order of 1 which means that it is probably
hopeless to consider the many-particle problem for graphene quite
rigorously. We will use the Thomas-Fermi theory\cite{lieb} which
is, actually, the simplest approximation in the density functional
approach. It is based on the two assumptions: (i) we neglect the
exchange-correlation potential in comparison with the Hartree
potential in Eq.(\ref{DF}) and (ii) we put $n\left(
\mathbf{r}^{\prime }\right) = n\left[ \mu -V\left(
\mathbf{r}^{\prime }\right) \right]$, $n\left( \mu \right) $ being
a density of homogeneous electron gas with chemical potential
$\mu$. The former assumption means that we are interested in the
long-wavelength response of the electron system and thus the
long-range Coulomb forces dominate over the local
exchange-correlation effects. The later one holds provided that
the external potential is smooth enough. A rigorous statement is
that an addition of \textit{constant} potential is equivalent to
the shift of the chemical potential. In particular, the
Thomas-Fermi theory gives an exact expression for static
inhomogeneous dielectric function $\epsilon(q)$ in the limit of
small wavevectors $q \rightarrow 0$\cite{pines,vons}. The
Thomas-Fermi theory of atoms is asymptotically exact in the limit
of infinite nuclear charge\cite{lieb}. Here we will use it just
for semiquantitative analysis of the problem.

In the Thomas-Fermi theory Eq.(\ref{DF}) reads
\begin{equation}
V_{ind}\left( \mathbf{r}\right) = \frac{e^{2}}{\epsilon }\int d\mathbf{r}%
^{\prime }\frac{n\left[ \mu -V\left( \mathbf{r}^{\prime }\right)
\right] -n\left( \mu \right) }{\left| \mathbf{r-r}^{\prime
}\right| }.  \label{ind}
\end{equation}

The function $n\left( \mu \right) $ is expressed via the density
of states $N\left( E\right) $:
\begin{equation}
n\left( \mu \right) =\int dEf\left( E\right) N\left( E\right)
=\int\limits^{\mu }dEN\left( E\right)   \label{nmu}
\end{equation}
where $f\left( E\right)$ is the Fermi function and the last
equality is valid for zero temperature (we will restrict ourselves
here only by this case). For the case of graphene with linear
energy spectrum near the crossing points $K$ and $K^{\prime }$ one
has
\begin{equation}
n\left( \mu \right) =\frac{1}{\pi }\frac{\mu \left| \mu
\right|}{\hbar ^{2}v_{F}^{2}}, \label{nmu1}
\end{equation}
where we have taken into account a factor 4 due to two valleys and
two spin projections.

Let us start first with the case of zero doping ($\mu =0$) where,
according to the linear response theory, there is no screening at
all. Substituting Eqs.(\ref{ind}), (\ref{impur}), and (\ref{nmu1})
into Eq.(\ref{total}), introducing the notation
\begin{equation}
V\left( r\right) =\frac{e^{2}}{\epsilon r}F\left( r\right)
\label{F}
\end{equation}
and integrating over the polar angle of vector $\mathbf{r}^{\prime
}$, we obtain a nonlinear integral equation for the function
$F\left( r\right):$
\begin{equation}
F\left( r\right) =Z-\frac{2Q}{\pi }\int\limits_{0}^{\infty
}\frac{dr^{\prime
}}{r^{\prime }}\frac{r}{r+r^{\prime }}K\left( \frac{2\sqrt{rr^{\prime }}}{%
r+r^{\prime }}\right) F^{2}\left( r^{\prime }\right)
\label{integral}
\end{equation}
where
\begin{equation}
K\left( k\right) =\int\limits_{0}^{\pi /2}\frac{d\varphi
}{\sqrt{1-k^{2}\sin ^{2}\varphi }}  \label{K}
\end{equation}
is the complete elliptic integral,
\begin{equation}
Q=2\left( \frac{e^{2}}{\epsilon \hbar v_{F}}\right) ^{2};
\label{Q}
\end{equation}
for the case of graphene on SiO$_{2}$ $Q\simeq 2$.

We will see below that, actually, the integral in right-hand side
of Eq.(\ref{integral}) is divergent at $r^{\prime }=0$; the reason
is that the expression (\ref {nmu1}) with the replacement of $\mu
$ by $V\left( r\right) $ is not applicable for a very small
distances when the potential becomes comparable with the
conduction bandwidth; thus we should introduce a cut-off at
$r^{\prime }\simeq a$ where $a$ is of order of a lattice constant.
An exact value of $a$ is not relevant, with a logarithmic
accuracy.

To proceed further we replace variables in Eq.(\ref{integral}),
$r^{\prime }=re^{t}, $ and introduce a notation
$\widetilde{F}\left( \ln r\right) =F\left( r\right).$ As a result,
Eq.(\ref{integral}) takes the form
\begin{equation}
\widetilde{F}\left( x\right) =Z-\frac{2Q}{\pi }\int\limits_{\ln a}^{x}dt%
\widetilde{F}^{2}\left( t\right) -\frac{2Q}{\pi
}\int\limits_{-\infty }^{\infty }dt\widetilde{F}^{2}\left(
x+t\right) \phi \left( t\right) , \label{Ftilda}
\end{equation}
where
\begin{equation}
\phi \left( t\right) =\frac{2}{\pi }\frac{K\left( \frac{1}{\cosh
t/2}\right) }{1+e^{t}}-\theta \left( -t\right),   \label{phi}
\end{equation}
$\theta \left( x>0\right) =1,\theta \left( x<0\right) =0.$ The
function $\phi \left( t\right) $ decays exponentially at
$t\rightarrow \pm \infty $ and has a logarithmic divergence at
$t=0$ (see Fig. 1). For large $x$ the last term in the right-hand
side of Eq.(\ref{Ftilda}) can be neglected. After that,
Eq.(\ref{Ftilda}) is transformed into a differential equation
which can be easily solved. As a result, we find the following
solution for the screening function $F$:
\begin{equation}
F\left( r\right) \simeq \frac{Z}{1+ZQ\ln \frac{r}{a}},
\label{solution}
\end{equation}
$r\gg a$.

\begin{figure}[tbp]
\includegraphics[width=7cm]{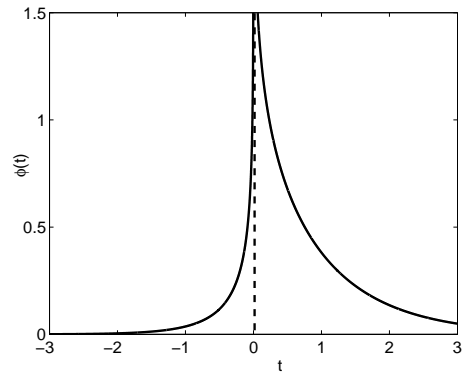}
\caption{Graph of the function $\phi\left( t\right)$
(Eq.(\ref{phi}))} \label{fig:1}
\end{figure}

This logarithmic screening of the Coulomb potential results from a
creation of electron-hole pairs in the vicinity of the impurity,
or, in terms of quantum electrodynamics (QED), a ``vacuum
polarization''\cite{blp,migdal}. This effect can be qualitatively
described in QED by an approach which is very similar to the
Thomas-Fermi theory used here\cite{migdal}.

As a result, at distances much larger than the lattice constant
charge-impurity potential in undoped graphene equals
\begin{equation}
V\left( r\right) \simeq \frac{e^{2}}{\epsilon r}\frac{1}{Q\ln
\frac{r}{a}}, \label{Vundoped}
\end{equation}
does not depend on the impurity charge $Z$ and is much weaker than
the bare potential $V_{0}\left( r\right) .$ This follows from the
fact that the ``effective fine-structure constant'' for graphene,
$\alpha$ is close to 1, instead of 1/137 in QED.

Consider now a generic case of doped graphene. In this case,
Eqs.(\ref{total}), (\ref{impur}), (\ref{ind}), and (\ref{nmu1})
result in the following integral equation for the total impurity
potential:
\begin{equation}
V\left( \mathbf{r}\right) +\frac{2 \left| \mu \right| e^{2}}{\pi
\epsilon
\hbar ^{2}v_{F}^{2}}\int d\mathbf{r}^{\prime }\frac{V\left( \mathbf{r}%
^{\prime }\right) }{\left| \mathbf{r-r}^{\prime }\right| }=\frac{Ze^{2}}{%
\epsilon r}-\frac{e^{2}}{\pi \epsilon \hbar ^{2}v_{F}^{2}}%
\int\limits_{r^{\prime }>a}d\mathbf{r}^{\prime }\frac{V^{2}\left( \mathbf{r}%
^{\prime }\right) }{\left| \mathbf{r-r}^{\prime }\right| }
\label{general}
\end{equation}
If one neglect the nonlinear term in the right-hand side of
Eq.(\ref{general}) this equation can be easily solved by the
Fourier transform; the result for the Fourier component of the
total potential, $v\left( q\right)$ reads\cite{nomura,ando}
\begin{equation}
v\left( q\right) =\frac{2\pi Ze^{2}}{\epsilon \left( q+\kappa
\right) } \label{fourier}
\end{equation}
where
\begin{equation}
\kappa =\frac{4e^{2} \left| \mu  \right|}{\epsilon \hbar
^{2}v_{F}^{2}} \label{kappa}
\end{equation}
is the inverse screening radius proportional to the Fermi wave
vector $k_{F}$. After inverse Fourier transformation one finds
\begin{equation}
V\left( r\right) =\frac{Ze^{2}}{\epsilon r}\left\{ 1-\frac{\pi \kappa r}{2}%
\left[ \mathbf{H}_{0}\left( \kappa r\right) -Y_{0}\left( \kappa r\right) %
\right] \right\}   \label{struve}
\end{equation}
with asymptotic behavior
\begin{equation}
V\left( r\right) \simeq \frac{Ze^{2}}{\epsilon r}\frac{1}{\left(
\kappa r\right) ^{2}}  \label{asymp}
\end{equation}
at $\kappa r\gg 1$; here $\mathbf{H}_{0}$ and $Y_{0}$ are Struve
and Neumann functions.

Estimating different terms in Eq.(\ref{general}) one can
demonstrate that the solution (\ref{solution}) is still correct
for $\kappa r\leq 1$ and the solution (\ref{struve}) - for $\kappa
r\gg 1,$ but with a replacement of $Z$ by
\begin{equation}
Z^{\ast }=Z-\frac{1}{\pi \hbar ^{2}v_{F}^{2}}%
\int\limits_{r^{\prime }>a}d\mathbf{r}^{\prime }V^{2}\left( \mathbf{r}%
^{\prime }\right) \label{eff}
\end{equation}
in the latter case. Analyzing contributions to the integral in the
right-hand side of Eq.(\ref{asymp}) from these two regions we
obtain our final result
\begin{equation}
Z^{\ast }\simeq \frac{Z}{1+ZQ\ln \frac{1}{\kappa a}}.
\label{final}
\end{equation}
This is the effective charge of impurity in graphene at distances
much larger than the lattice constant. Since we always have
$k_{F}a\ll 1$ this means that it is the charge that determines
electron scattering by a long-range part of charge impurity
potential in graphene. This weakens essentially this scattering
mechanism since $Q\ln \frac{1}{\kappa a}$ is of order of ten for
typical charge carrier concentrations. Perturbative estimations of
the electron mobility\cite{nomura} should be thus multiplied by
this factor squared. As a result, the mobility for the same
parameters turns out to be two order of magnitude larger. Instead
of concentration-independent mobility, we obtain a mobility
proportional to $\ln ^{2}\left( k_{F}a\right)$. This weak
dependence on the charge-carrier concentration is probably
consistent with the experimental data\cite{private}. More
accurately, one should use an expression for the mobility obtained
by Ando\cite{ando} (see Eq.(3.27) and Fig. 5 of that paper) but
with the replacement of $Z$ by $Z^{\ast }$ when calculating the
strength of the Coulomb interaction.

I am thankful to Andre Geim and Kostya Novoselov for valuable
discussions stimulating this work.

\end{document}